\documentclass[conference]{IEEEtran}
\IEEEoverridecommandlockouts
\usepackage{cite}
\usepackage{amsmath,amssymb,amsfonts}
\usepackage{algorithmic}
\usepackage{graphicx}
\usepackage{textcomp}
\usepackage{xcolor}
\usepackage{multirow}
\usepackage{caption}
\usepackage{subcaption}
\usepackage{svg}
\def\BibTeX{{\rm B\kern-.05em{\sc i\kern-.025em b}\kern-.08em
    T\kern-.1667em\lower.7ex\hbox{E}\kern-.125emX}}

\usepackage{tikz}

\newcommand\copyrighttext{%
  \footnotesize \textcopyright 2024 IEEE. Personal use of this material is permitted.  Permission from IEEE must be obtained for all other uses, in any current or future media, including reprinting/republishing this material for advertising or promotional purposes, creating new collective works, for resale or redistribution to servers or lists, or reuse of any copyrighted component of this work in other works.}
\newcommand\copyrightnotice{%
\begin{tikzpicture}[remember picture,overlay]
\node[anchor=south,yshift=10pt] at (current page.south) 
  {\fbox{\parbox{\dimexpr\textwidth-\fboxsep-\fboxrule\relax}{\copyrighttext}}};
\end{tikzpicture}%
}

\begin{document}

\title{Initial Characterization of Healthy and Malignant \textit{in vivo} and \textit{ex vivo} Human Colon Tissues under Surgery Procedures\\
\thanks{This work was funded by the eSCOPE project funded by Hospital La Fe and Universitat Politècnica de Valencia, and by Program Investigo (INVEST/2022/199), held by the Conselleria de Innovación, Universidades, Investigación y Sociedad Digital de la Generalitat Valenciana, funded by Next Generation EU within the framework Plan de Recuperación, Transformación y Resiliencia de España.

This work has been additionally funded by Ramón y Cajal fellowship RYC2021-033207-I, by EMERGE-5G project PID2020-115005RJ-I00, and by HEROE-6G CNS2023-143570 project, all funded by Agencia Estatal de Investigación MCIN/AEI/10.13039/501100011033 and by European Union NextGenerationEU/PRTR.
}
}

\author{\IEEEauthorblockN{Sergio Micó-Rosa}
\IEEEauthorblockA{\textit{iTEAM} \\
\textit{Universitat Politècnica de València}\\
Valencia, Spain \\
sermiro@iteam.upv.es}
\and
\IEEEauthorblockN{Concepcion Garcia-Pardo}
\IEEEauthorblockA{\textit{iTEAM} \\
\textit{Universitat Politècnica de València}\\
Valencia, Spain \\
cgpardo@iteam.upv.es}
\and
\IEEEauthorblockN{Matteo Frasson}
\IEEEauthorblockA{\textit{Digestive Diseases Area} \\
\textit{Hospital Universitari i Poltècnic La Fe}\\
Valencia, Spain \\
dr.frasson.matteo@gmail.com}
\and
\IEEEauthorblockN{Narcís Cardona}
\IEEEauthorblockA{\textit{iTEAM} \\
\textit{Universitat Politècnica de València}\\
Valencia, Spain \\
ncardona@iteam.upv.es}
\and
\IEEEauthorblockN{Vicente Pons-Beltrán}
\IEEEauthorblockA{\textit{Digestive Diseases Area} \\
\textit{Hospital Universitari i Poltècnic La Fe}\\
Valencia, Spain \\
pons\_vicbel@gva.es}
\and
\IEEEauthorblockN{Pedro López-Muñoz}
\IEEEauthorblockA{\textit{Digestive Diseases Area} \\
\textit{Hospital Universitari i Poltècnic La Fe}\\
Valencia, Spain \\
pedro.lopez8928@gmail.com}
}

\maketitle
\copyrightnotice
\begin{abstract}
The dielectric characterization of human tissues can play a crucial role in the development of new medical diagnostic tools. In particular, the characterization of healthy and pathological tissues can provide vital information for diagnosis. In this paper, preliminary results from a small-scale measurement campaign conducted in 0.5-26.5 GHz during real surgeries on healthy and malignant human colon tissues are presented. Those measurements were carried out externally to the colon, without direct contact to the tumor growing inside the colon. Furthermore, different tumor stages are taken into account. Initial findings reveal that advanced tumor stages are related with increased higher values of dielectric properties in malignant tumor tissues compared to the healthy ones.
\end{abstract}

\begin{IEEEkeywords}
Dielectric properties, open-ended coaxial, tissue characterization, tumor stage
\end{IEEEkeywords}

\section{Introduction}
The dielectric characterization of human tissues plays a crucial role in various applications, ranging from the development of in-body sensors and electronic devices to the assessment of Specific Absorption Rate (SAR) and electromagnetic exposure, as well as applications in electromagnetic simulations and medical diagnostic tools. Especially in the field of medical diagnostics, achieving a detailed characterisation of tissues is essential in order to identify discernible patterns that can serve as distinctive parameters in the diagnostic process.

Numerous studies have explored the dielectric properties of tissues, and the Gabriel’s database \cite{b1} has established itself as the main reference, containing characterisations of more than 50 tissues. In particular, some studies have highlighted the higher permittivity observed in malignant tissues compared to healthy ones \cite{b2,b3,b4}. A subset of these studies focuses on colon cancer, attempting to clarify the differences between malignant and healthy tissue \cite{b5,b6,b7}. However, most of these investigations are carried out \textit{ex vivo}, after removal of the tumor from patients.

This article focuses on determine how tumor stage influences the permittivity measured during tissue characterization, especially considering that measurements are performed externally in the colon while the tumor develops inside the colon. To approach this objective, a preliminary study examining the relationship between tumor stage and the difference in dielectric properties between cancerous and healthy tissue within the frequency range of 0.5 to 26.5 GHz is presented. In particular, measurements were performed both \textit{ex vivo} and \textit{in vivo} during real surgeries.

The organization of this paper is structured as follows. Section II details the measurement system and methodology employed, while Section III presents the results. The conclusion and future work are outlined in Section IV.

\section{Methodology}

\subsection{Measurement system}
\begin{figure}[t]
    \centerline{\includegraphics[scale=1.1]{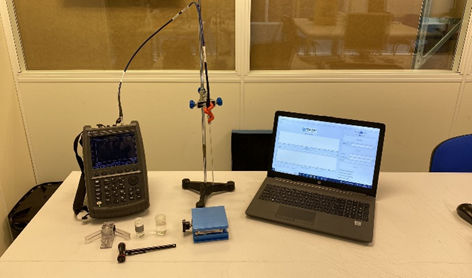}}
    \caption{Open-ended coaxial proble system: VNA (left),open-ended coaxial probe (center) and laptop (right).}
    \label{fig:meas_setup}
\end{figure}

The measurements in this study were performed using the open-ended coaxial probe technique. This method is widely used for tissue characterization due to its many strengths such as its wide bandwidth, its high measurement speed, its suitability for semi-solid materials and the fact that it is a non-destructive technique. The setup is based on a Vector Network Analyzer (VNA, model Keysight FieldFox N9918A), an open-ended coaxial probe (Keysight N1501A), a flexible coaxial cable to connect the VNA with the probe and a laptop to control the system and process the measurements. This setup is shown in Fig.~\ref{fig:meas_setup}.

The principle of operation is as follows: the VNA sends a signal to the probe, where part of it is reflected by the tissue in contact with the probe and this reflection is captured by the VNA. The laptop then processes it to obtain the permittivity of the tissue from this reflection. For this, a previous calibration of the system is necessary, which is carried out by means of 4 elements with known properties (air, short circuit, water and methanol), as described in \cite{b8}.

The configuration parameters used in the setup are listed in Table~\ref{tab1}.

\begin{table}[b]
\caption{Configuration Parameters of the Measurement System}
\begin{center}
\begin{tabular}{|c|c|c|c|c|}
\hline
\textbf{Frequency}&\multirow{ 2}{*}{\textbf{Frequency points}}&\textbf{IF BW}&\textbf{Power}&\multirow{ 2}{*}{\textbf{Sweeps}} \\
\textbf{(GHz)}& &\textbf{(kHz)}&\textbf{(dBm)}& \\
\hline 
\multirow{ 2}{*}{0.5 - 26.5}&1601 (\textit{ex vivo})&\multirow{ 2}{*}{3}&\multirow{ 2}{*}{-15}&\multirow{ 2}{*}{3} \\
 &401 (\textit{in vivo})& & & \\
\hline
\end{tabular}
\label{tab1}
\end{center}
\end{table}

\subsection{Measurement procedure}
The measurements were carried out at surgery procedures under different conditions at the Hospital La Fe in Valencia. The study was previously approved by the ethics committee of the Hospital la Fe (ref. 2020-483-1) and of the Univ. Politècnica de València (ref. P08\_26\_11\_20).

In this measurement campaign, the different stages of patient’s tumor were taking into account. The tumor stage referred to is T1-T4, which describes the size and location of the tumor, on a scale from 1 to 4 \cite{b9}. The larger the tumor size or the deeper the growth into the different layers of the intestine, the higher the number. In the present study there are measurements of advanced tumor stages. In particular, stages T3, T4a and T4b, ordered from lowest to highest stage.

\begin{figure}[htbp]
    \centerline{\includegraphics[scale=0.45]{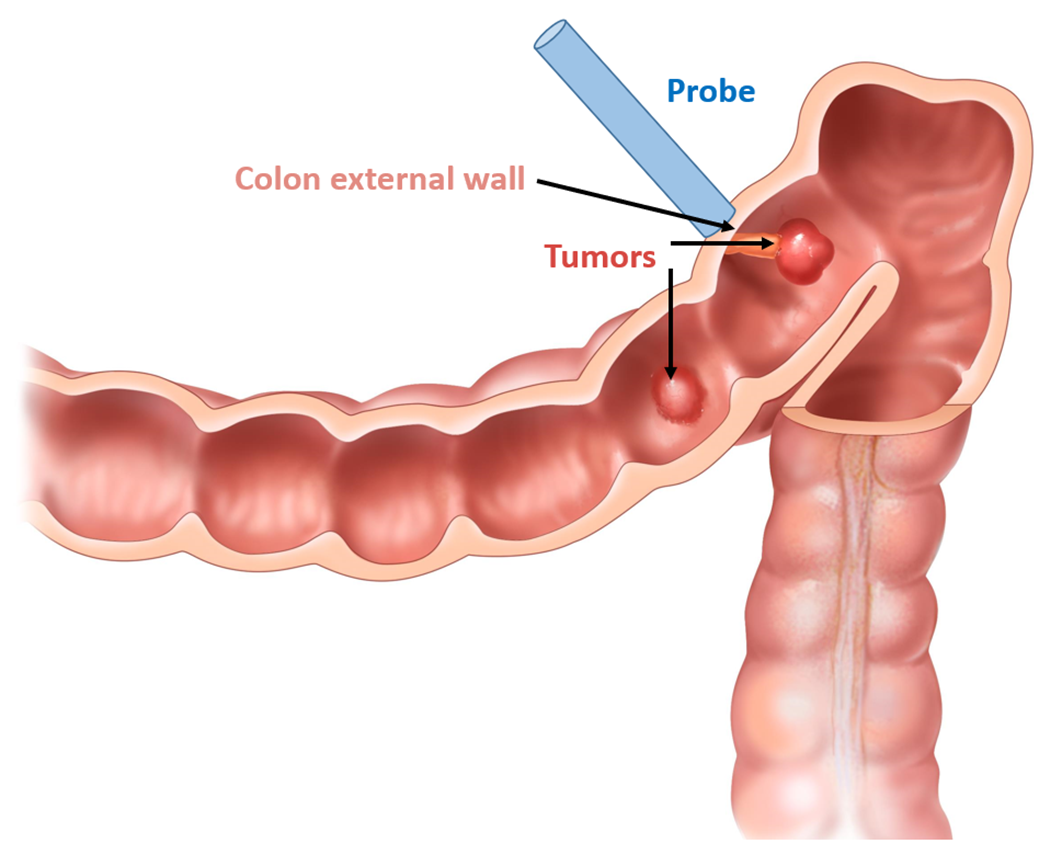}}
    \caption{Recreation of measurement carried out in the external wall of the colon.}
    \label{fig:meas_recreation}
\end{figure}

\begin{figure}[htbp]
    \centerline{\includegraphics[scale=1.1]{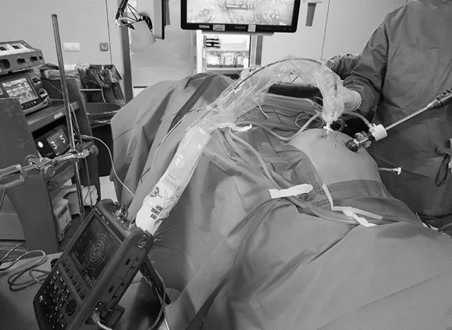}}
    \caption{Image of the measurement system in an \textit{in vivo} situation in a surgery. It shows how the cable is protected with a sterile bag to protect the patient (center) and how the probe is inserted through an incision made for the surgery (right).}
    \label{fig:invivo_meas}
\end{figure}

\begin{figure}[htbp]
    \centerline{\includegraphics[scale=0.53]{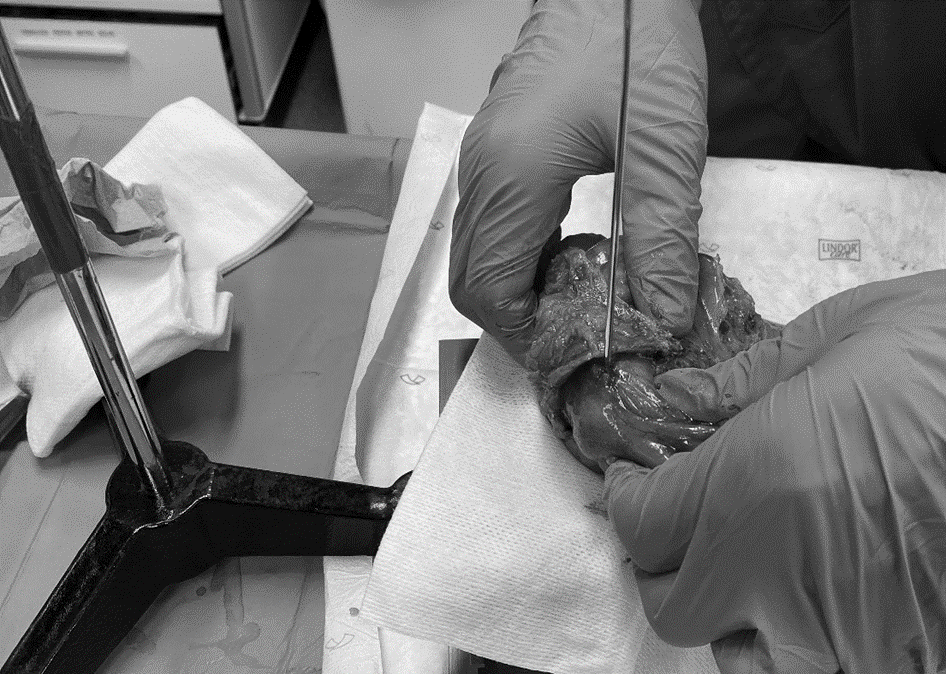}}
    \caption{Image of the measurement system in an \textit{ex vivo} situation in a surgery. It shows how the tissue is moved close to the calibrated probe with the help of an small elevator and by hands to make sure good contact between the tissue and the probe.}
    \label{fig:exvivo_meas}
\end{figure}

\begin{itemize}
    \item \textit{In vivo}: Once the patient is anaesthetised to proceed with the surgery, surgeons access the colon through an incision in the abdominal area either by laparotomy or laparoscopy. Thus, when measuring, the surgeon accesses the tissues of the organs from the outside of them, so the outer layers of the tissues are measured (see Fig.~\ref{fig:meas_recreation}).
    
    As the measurement is performed during surgery, the probe must be previously sterilised when entering in contact to the patient’s tissue. It is therefore impossible to calibrate the system beforehand, so a pre-set calibration is used. Thus, during surgery, it is the surgeon who places the probe in the tissues to be measured while the system is controlled externally from a laptop located on a table near the patient, as shown in Fig.~\ref{fig:invivo_meas}. Two different locations of cancer and healthy colon tissues are measured. Once the measurements are completed, a post-calibration is performed to post-process the data properly. Meanwhile, surgeons begin operating on the patient to remove the tumor. A total of 27 \textit{in vivo} measurements are taken from 7 patients, of those there are 1 patient with tumor stage T4b, 2 with T4a and 4 with T3.
    \item \textit{Ex vivo}: Just before the removal of the piece by the surgeons, the configuration and calibration of the system is performed so the measurement can be started as soon as the piece is extracted. In the meantime, surgeons remove the piece of the tumour and it is placed on a surgical table in the operating theatre anteroom where the equipment is ready to be measured. Then, surgeons explain where the tumor should be measured and finally the measurements of both the tumor and healthy colon are carried out. Since the extracted sample is of considerable size, as the tumor and part of the surrounding colon (approximately 10 cm around) are removed, it is not possible to measure inside the colon, so the measurement is performed on the external side of the colon (see Fig.~\ref{fig:exvivo_meas}). Two different locations are measured for the cancer and healthy colon tissues. There are a total of 37 \textit{ex vivo} measurements of 8 patients, of which there are 1 patient with tumor stage T4b, 2 with T4a and 5 with T3.
\end{itemize}

\subsection{Data processing}\label{sec:2C}
To process the data obtained from the surgeries, the mean value of the sweeps of each tissue is performed. Then, they were averaged by tissue status (healthy or tumor) and measurement scenario (\textit{ex vivo} or \textit{in vivo}). In addition, in the case of the study by tumor stage, the same analysis was carried out for each stage separately, grouping the patients' tissues by tumor stage. Data on tumor stages are provided by the hospital's anatomical pathology unit.

As is well known, permittivity is a frequency-dependent complex value which is referred to as relative to the vacuum permittivity, as shown in \eqref{eq1}.

\begin{equation}
\varepsilon_r(f)=\varepsilon(f)/\varepsilon_0 = \varepsilon_r^{'}(f)-j\varepsilon_r^{''}(f)\label{eq1}
\end{equation}

where $f$ is the frequency, $\varepsilon_r$ is the complex value of the relative permittivity, $\varepsilon$ is the absolute permittivity and $\varepsilon_0$ is the permittivity of the vacuum. Thus, $\varepsilon_r^{'}$ is the real part of the relative permittivity, also known as the dielectric constant, and $\varepsilon_r^{''}$ is the imaginary part, known as the loss factor.

The measured permittivity is obtained as described in \cite{b8}. For each measurement point, 3 sweeps are obtained and averaged. Thus, for each measuring point of each tissue a measured permittivity value is obtained.

In addition, the measured permittivity can be modelled. The purpose of this is to obtain a curve without the noise or oscillations that the original signal could present. In this case, the mean of the measured permittivity data is fitted with a Cole-Cole equation:

\begin{equation}
\varepsilon_r(f)=\varepsilon_{\infty}+\sum_{m=1}^{M}\frac{\Delta\varepsilon_m}{1+(j\omega\tau_m)^{1-\alpha_m}}+\frac{\sigma_s}{j\omega\varepsilon_0}\label{eq2}
\end{equation}

where $M$ is the number of poles of the Cole-Cole equation and each one represents one dispersion area, $\omega$ is the angular frequency, $f$ is the frequency and $\Delta\varepsilon_m$, $\tau_m$ and $\sigma_s$ are the coefficients to fit. For this study, 2 poles were used because they were sufficient to correctly model the permittivity of the measurements. Using a larger number of poles did not significantly increase the accuracy of the model.

Another approach to analyze the data is to look at the difference between the measured permittivity of tumor and healthy tissue. It is calculated as shown in \eqref{eq3} and \eqref{eq4} depending on the dielectric constant or loss factor; where the average value of the measured permittivity of each tissue is used to calculate the difference.

\begin{equation}
\Delta\varepsilon_r^{'}(f)=\overline{\varepsilon_r^{'}}_{tumor}(f)-\overline{\varepsilon_r^{'}}_{healthy}(f) \label{eq3}
\end{equation}

\begin{equation}
\Delta\varepsilon_r^{''}(f)=\overline{\varepsilon_r^{''}}_{tumor}(f)-\overline{\varepsilon_r^{''}}_{healthy}(f) \label{eq4}
\end{equation}

This difference is calculated for each patient separately and fitted with a cubic equation. Since each patient may have more than one measurement of tumor or healthy tissue, the average of these is used to give a single difference per patient so that each one has the same weight when calculating the average difference. Then, by grouping these differences according to \textit{in vivo} or \textit{ex vivo} or their tumor stage, they can be analysed separately by groups.

\section{Results}
\begin{figure}[htbp]
    \begin{subfigure}[b]{0.46\textwidth}
        \centering\includegraphics[trim=30 0 45 22,clip,width=\textwidth]{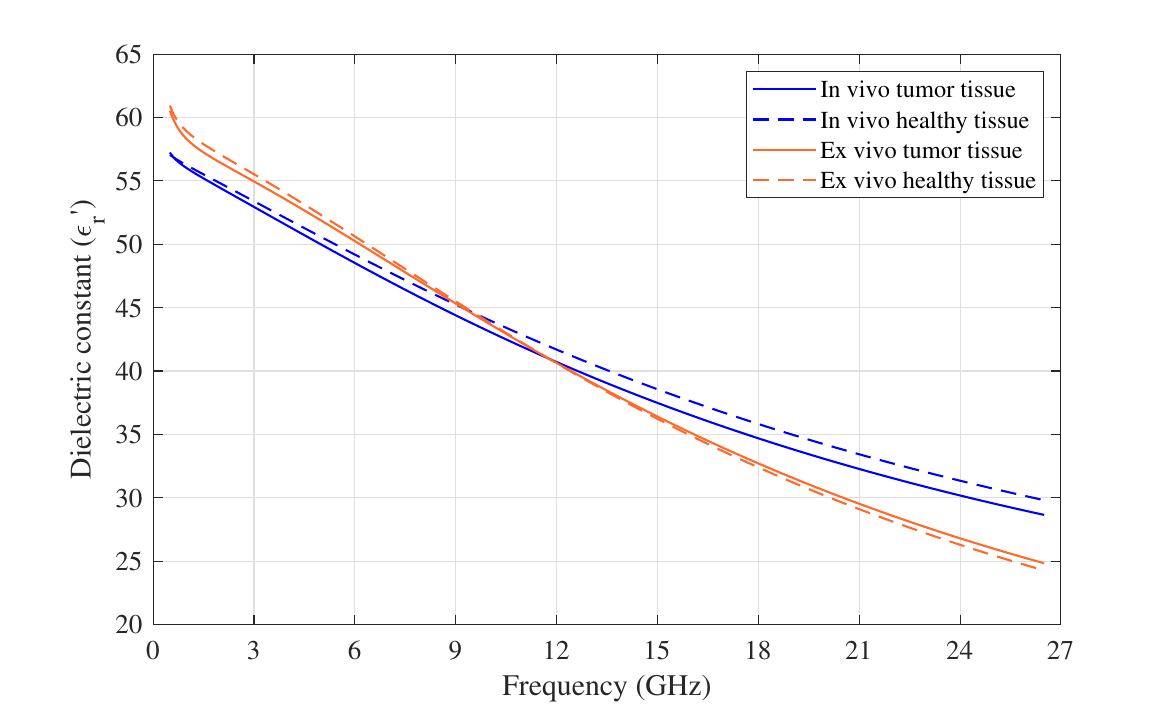}
        \caption{Dielectric constant.}
        \label{fig:DC_vs_freq}
    \end{subfigure}
    
    \begin{subfigure}[b]{0.46\textwidth}
        \centering\includegraphics[trim=30 0 45 22,clip,width=\textwidth]{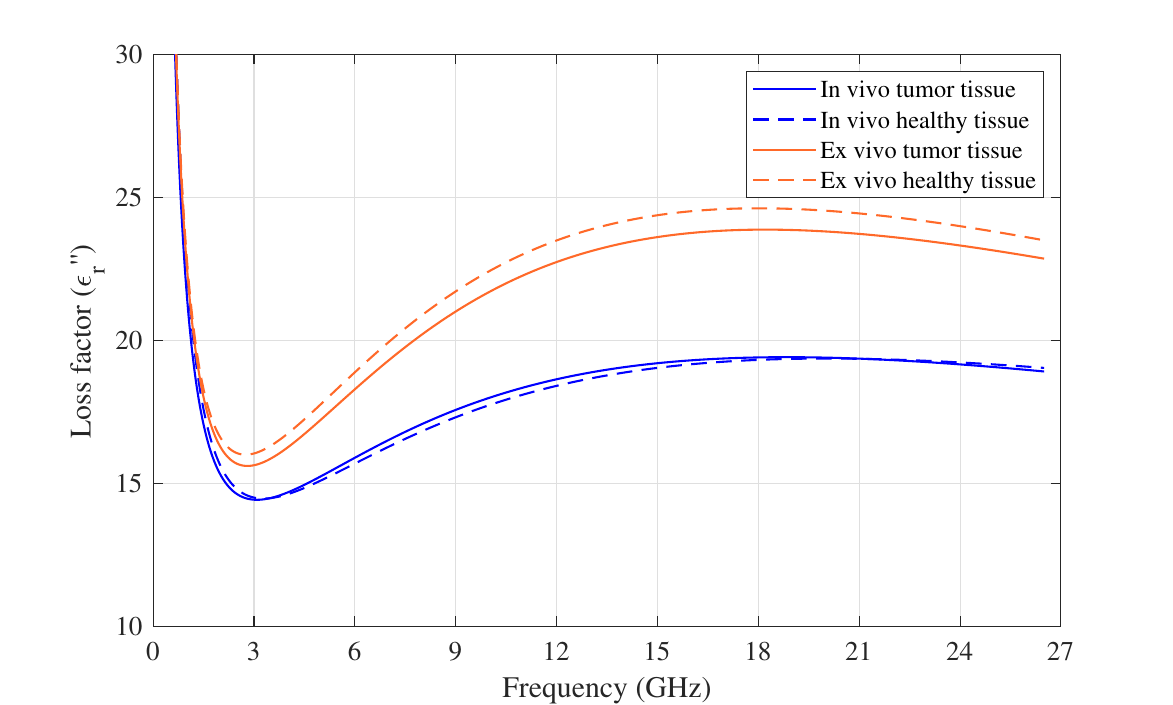}
        \caption{Loss factor.}
        \label{fig:LF_vs_freq}
    \end{subfigure}
    \caption{Means of the dielectric constant and loss factor of healthy and tumor tissues in \textit{ex vivo} and \textit{in vivo} measurements.}
    \label{fig:DC&LF_vs_freq}
\end{figure}

The Cole-Cole model (according to \eqref{eq2}) of the mean of the dielectric properties of the colon measured on both tumor and healthy tissue at \textit{ex vivo} and \textit{in vivo} scenarios are presented in Fig.~\ref{fig:DC&LF_vs_freq}, where all the different tumor stages are not being taken into account so they are grouped together as tumor tissue.

As can be noticed, there is practically no difference between healthy and tumor tissues both \textit{ex vivo} and \textit{in vivo}. This differs from the literature \cite{b5,b6,b7}, where tumor tissue has a higher permittivity value, both in real and imaginary part, than healthy tissue. This is due to the fact that the measures in this campaign are performed externally to the colon, as opposed to previous studies where the healthy and tumor tissue was accessed directly from the inside. Also, all the different stages are grouped together.

Table~\ref{tab2} shows the fitted mean values of the difference between tumor and healthy tissue for various frequencies both \textit{ex vivo} and \textit{in vivo} (according to \eqref{eq3} and \eqref{eq4}). Where each patient has the same weight when averaging as explained in Section~\ref{sec:2C}.

\begin{table}[b]
\caption{Difference in Dielectric Constant and Loss Factor at various Frequencies for \textit{Ex vivo} and \textit{In vivo} Measurements}
\begin{center}
\begin{tabular}{|c|c|c|c|c|c|c|}
\hline
\textbf{Scenario}&\multicolumn{2}{|c|}{\textbf{2.45 GHz}}&\multicolumn{2}{|c|}{\textbf{12.5 GHz}}&\multicolumn{2}{|c|}{\textbf{18 GHz}} \\
\cline{2-7} 
 &\(\Delta\varepsilon_r^{'}\)&\(\Delta\varepsilon_r^{''}\)&\(\Delta\varepsilon_r^{'}\)&\(\Delta\varepsilon_r^{''}\)&\(\Delta\varepsilon_r^{'}\)&\(\Delta\varepsilon_r^{''}\) \\
\hline
\textit{Ex vivo}& -0.97& -0.38& -0.23& -0.83& 0.06& -0.86  \\
\hline
\textit{In vivo}& 0.96& 0.27& 0.04& 0.76& -0.26& 0.64  \\
\hline
\end{tabular}
\label{tab2}
\end{center}
\end{table}

It shows that there is no difference between malignant and healthy tissue for each patient on average. It should be highlighted that in this case, measurements were performed in the outer part of the colon, while the tumor grows from inside the colon, contrary to what has been reported in the literature. Thus, tumors in more advanced stages must have gone through more layers of the colon and should therefore be more perceptible for measurements taken from the outside. This is because the penetration depth of the measurement system is not very high, so any intermediate layer between the tumor and the probe would significantly affect the measurement result \cite{b10}.

\begin{figure}[t]
    \begin{subfigure}[h]{0.46\textwidth}
        \centering\includegraphics[trim=30 0 45 22,clip,width=\textwidth]{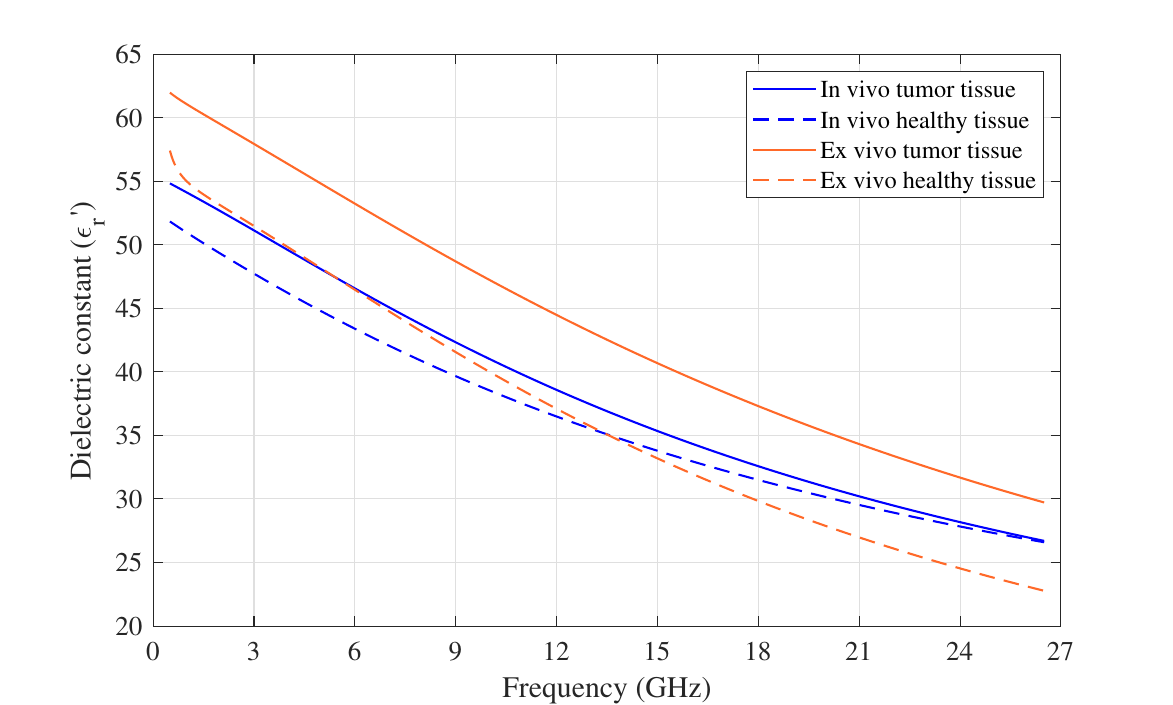}
        \caption{Dielectric constant.}
        \label{fig:DC4b_vs_freq}
    \end{subfigure}
    
    \begin{subfigure}[h]{0.46\textwidth}
        \centering\includegraphics[trim=30 0 45 22,clip,width=\textwidth]{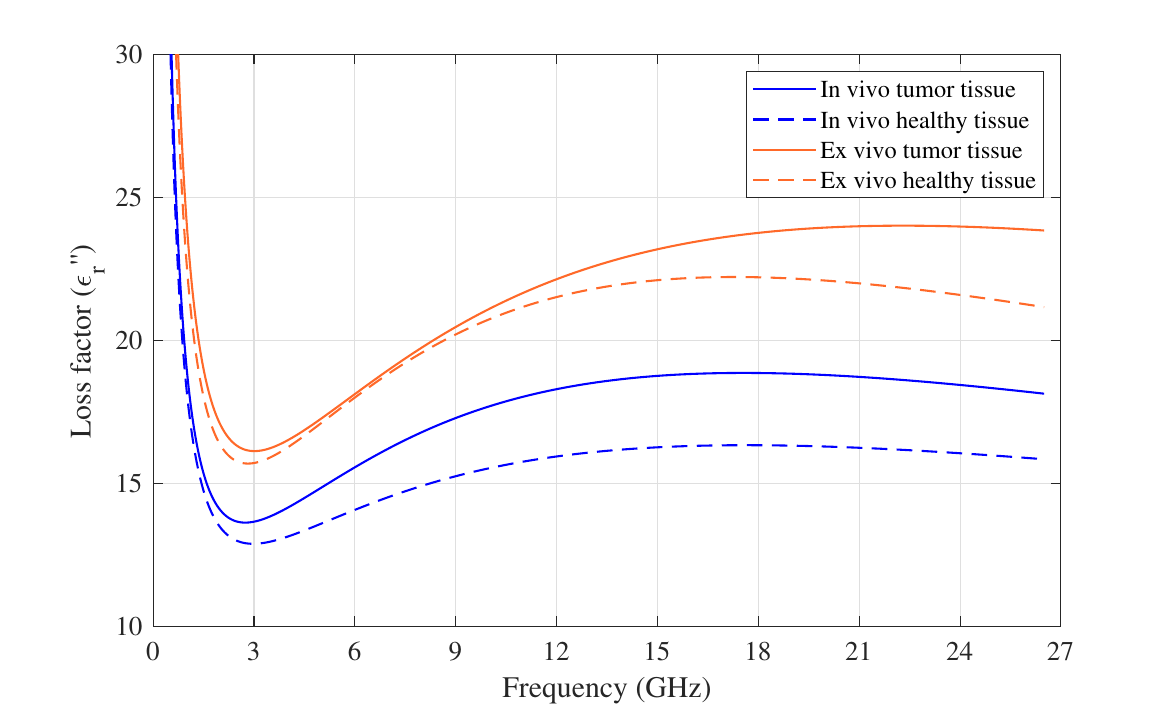}
        \caption{Loss factor.}
        \label{fig:LF4b_vs_freq}
    \end{subfigure}
    \caption{Means of the dielectric constant and loss factor of healthy and tumor tissues for tumor stage T4b in \textit{ex vivo} and \textit{in vivo} measurements.}
    \label{fig:DC&LF4b_vs_freq}
\end{figure}

\begin{figure}[t]
    \begin{subfigure}[h]{0.46\textwidth}
        \centering\includegraphics[trim=30 0 45 22,clip,width=\textwidth]{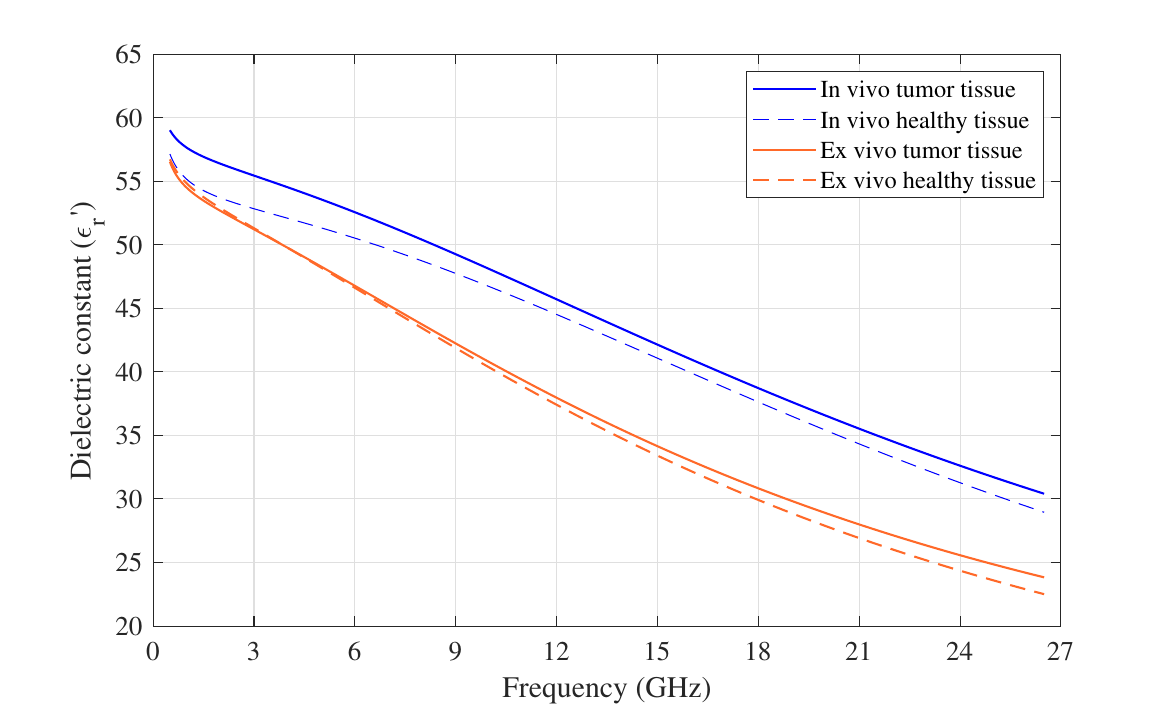}
        \caption{Dielectric constant.}
        \label{fig:DC4a_vs_freq}
    \end{subfigure}
    
    \begin{subfigure}[h]{0.46\textwidth}
        \centering\includegraphics[trim=30 0 45 22,clip,width=\textwidth]{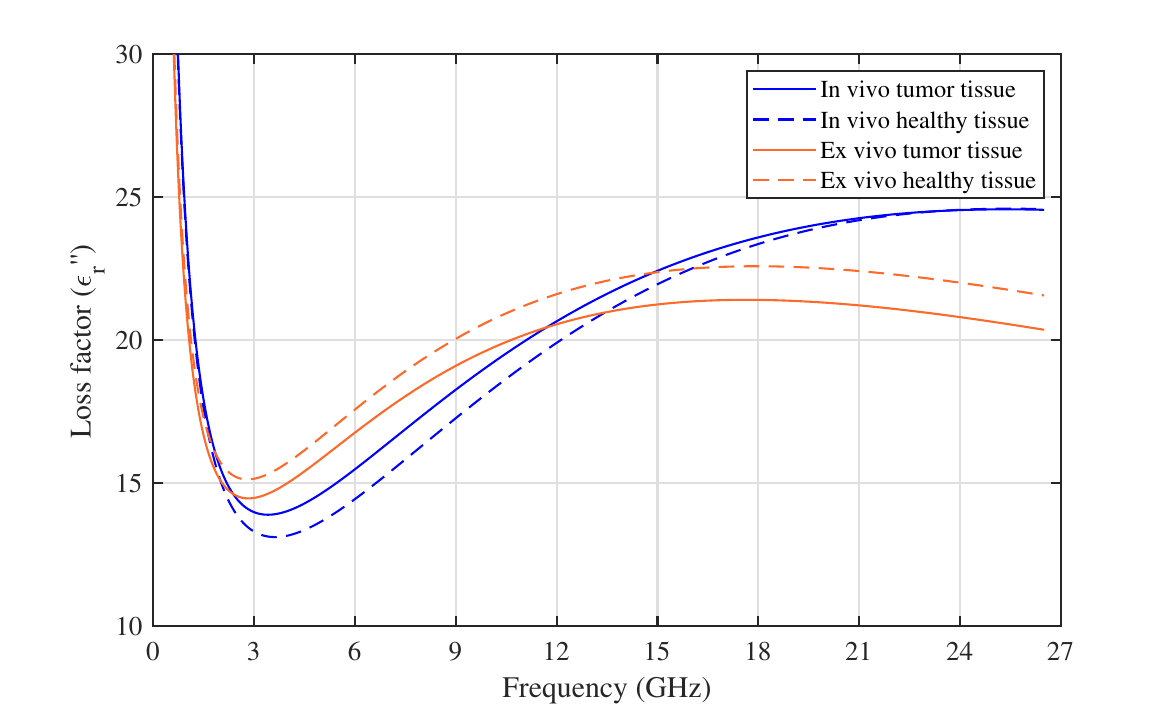}
        \caption{Loss factor.}
        \label{fig:LF4a_vs_freq}
    \end{subfigure}
    \caption{Means of the dielectric constant and loss factor of healthy and tumor tissues for tumor stage T4a in \textit{ex vivo} and \textit{in vivo} measurements.}
    \label{fig:DC&LF4a_vs_freq}
\end{figure}

\begin{figure}[t]
    \begin{subfigure}[h]{0.46\textwidth}
        \centering\includegraphics[trim=30 0 45 22,clip,width=\textwidth]{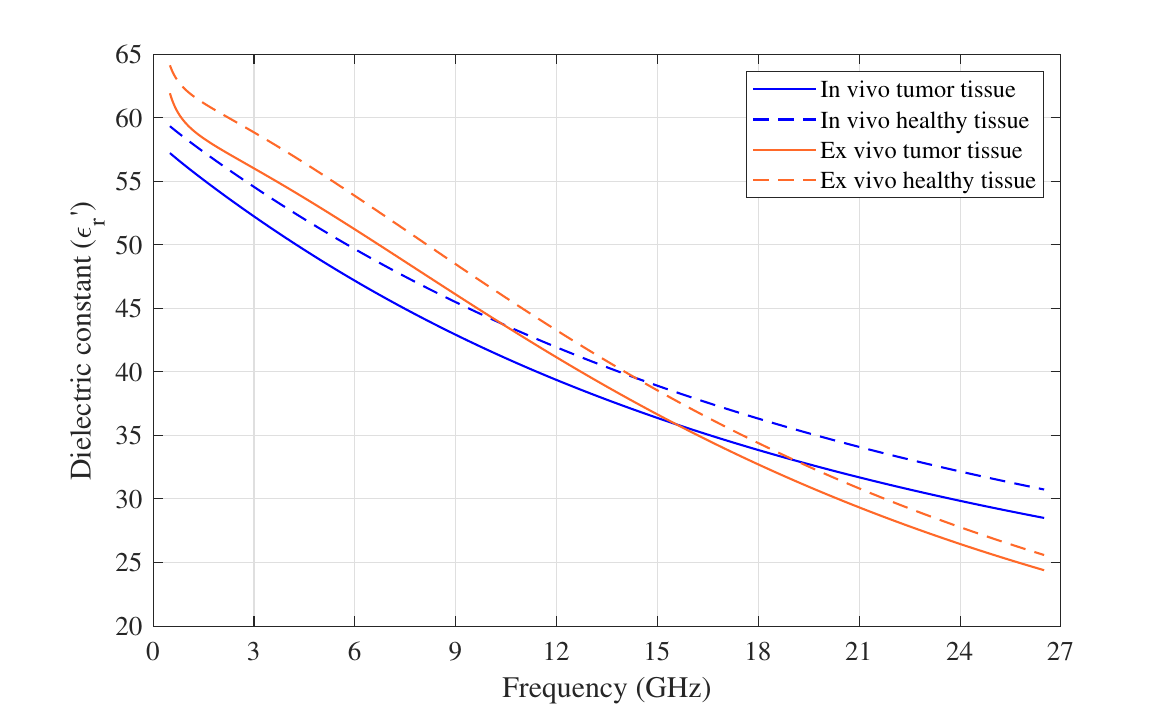}
        \caption{Dielectric constant.}
        \label{fig:DC3_vs_freq}
    \end{subfigure}
    
    \begin{subfigure}[h]{0.46\textwidth}
        \centering\includegraphics[trim=30 0 45 22,clip,width=\textwidth]{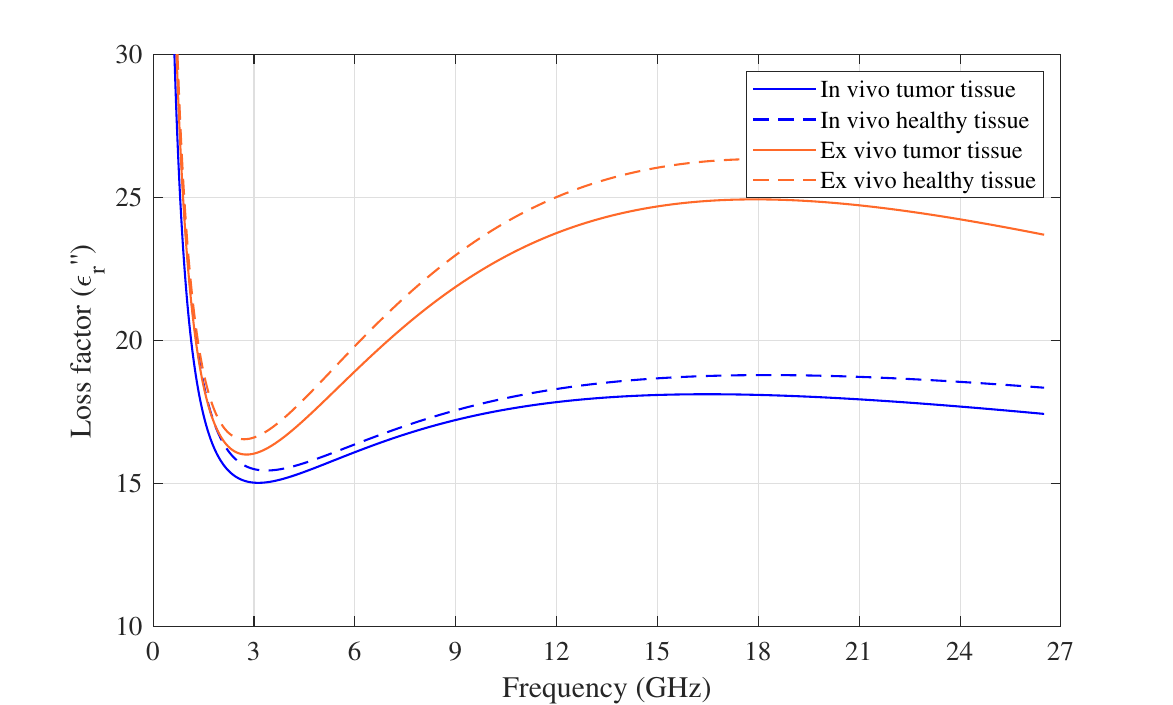}
        \caption{Loss factor.}
        \label{fig:LF3_vs_freq}
    \end{subfigure}
    \caption{Means of the dielectric constant and loss factor of healthy and tumor tissues for tumor stage T3 in \textit{ex vivo} and \textit{in vivo} measurements.}
    \label{fig:DC&LF3_vs_freq}
\end{figure}

Therefore, grouping the patients by their respective tumor stages, the Cole-Cole model of the mean of tumor and healthy tissue for the different groups for \textit{ex vivo} and \textit{in vivo} scenarios was calculated, as shown in Fig.~\ref{fig:DC&LF4b_vs_freq},~\ref{fig:DC&LF4a_vs_freq} and~\ref{fig:DC&LF3_vs_freq}.

As can be seen in the figures, only tumor stage T4b shows a considerable difference between tumor and healthy tissue for both \textit{ex vivo} and \textit{in vivo} scenarios in both the dielectric constant and the loss factor depending on the frequency. The same occurs for stage T4a but only for the \textit{in vivo} situation, while for stage T3 it does not occur in any case. It may be mentioned that the differences in each stage between \textit{ex vivo} and \textit{in vivo} should not be taken into account since different patients were measured.

In addition, the fitted mean of the differences between tumor and healthy tissue in each group was calculated separately, giving the results showed at Table~\ref{tab3} for different frequencies considering also \textit{ex vivo} and \textit{in vivo} scenarios.

\begin{table}[b]
\caption{Difference in Dielectric Constant and Loss Factor at various Frequencies Depending on Tumor Stage for Ex vivo and In vivo Measurements}
\begin{center}
\begin{tabular}{|c|c|c|c|c|c|c|c|}
\hline
\multirow{ 2}{*}{\textbf{Scenario}}&\textbf{Tumor}&\multicolumn{2}{|c|}{\textbf{2.45 GHz}}&\multicolumn{2}{|c|}{\textbf{12.5 GHz}}&\multicolumn{2}{|c|}{\textbf{18 GHz}} \\
\cline{3-8} 
 &\textbf{stage}&\(\Delta\varepsilon_r^{'}\)&\(\Delta\varepsilon_r^{''}\)&\(\Delta\varepsilon_r^{'}\)&\(\Delta\varepsilon_r^{''}\)&\(\Delta\varepsilon_r^{'}\)&\(\Delta\varepsilon_r^{''}\) \\
\hline
\multirow{ 3}{*}{\textit{Ex vivo}}& T3& -2.61& -0.44& -1.84& -1.14& -1.47& -1.28 \\
\cline{2-8}
 & T4a& -0.40& -0.67& 0.06& -0.88& 0.22& -1.12 \\
\cline{2-8}
 & T4b& 6.10& 0.53& 7.25& 0.82& 7.38& 1.75 \\
\hline
\multirow{ 3}{*}{\textit{In vivo}}& T3& -1.15& -0.08& -1.53& -0.03& -1.39& -0.00 \\
\cline{2-8}
 & T4a& 3.98& 0.67& 2.22& 1.28& 1.29& 0.90 \\
\cline{2-8}
 & T4b& 3.35& 0.91& 1.97& 2.27& 1.16& 2.66 \\
\hline
\end{tabular}
\label{tab3}
\end{center}
\end{table}

It is remarkable how larger differences between malignant and heathy colon tissues have been obtained for more advanced tumor stages in both real and imaginary parts in either \textit{ex vivo} or \textit{in vivo}. Only stage T4b shows positive values in both real and imaginary parts in \textit{ex vivo} and \textit{in vivo}. T4a \textit{in vivo} shows high values for the dielectric constant difference, reaching a value of 3.98 for the ISM frequency of 2.45 GHz. However, the highest differences are obtained for the T4b stage in \textit{ex vivo} scenario for the real part, with a maximum value of 7.38 at 18 GHz, and in T4b \textit{in vivo} for the imaginary part reaching 2.6 at 18 GHz. It is clearly observed that for the T3 stage, positive values are never reached, neither in dielectric constant nor in loss factor, being only in the \textit{in vivo} case where values of approximately 0 in the loss factor are obtained.

These results show how more advanced tumor stages show greater differences between tumor and healthy tissue. This must be due to the fact that more advanced stages grow more towards the outer layers of the colon, so it is easier to actually measure the tumor when the measurement is taken from the outside of the colon. Because, as mentioned above, the penetration depth of the probe is not very deep. Thus, intermediate layers between the tumor and the point on the outer layer of the colon, where the probe is placed to measure, can make the tumor appear to be hidden from the probe.

\section{Conclusions and future work}
This study presents a preliminary characterization of healthy and malignant human colon tissues, both \textit{ex vivo} and \textit{in vivo} during surgery procedures, focusing on the relationship between permittivity differences and tumor stage. Advanced tumor stages show greater permittivity disparities, probably due to their greater growth to the external layers of the colon, where the measurements are carried out by measuring outside. These findings, in the range of 0.5-26.5 GHz, are promising but considered preliminary with data from 7 \textit{in vivo} and 8 \textit{ex vivo} surgeries.

It should be noted that the measurement protocol have been modified so that they can be used in real surgeries and thus be approved by ethics committees. Therefore, \textit{ex vivo} and \textit{in vivo} measurements are performed differently, with \textit{in vivo} being the most novel, where a predefined calibration and post-calibration of the results is necessary due to the use of a sterile probe during the surgery.

The novelty of these results suggests a possible path for future research in medical diagnostics. Future efforts should extend the measurement campaign, comparing the results of external and internal colon measurements, possible only in \textit{ex vivo} scenario. This could validate the hypothesis that early stage tumors are only detectable internally, while advanced stages may be measurable both externally and internally.

\bibliographystyle{IEEEtran} 
\bibliography{refs} 

\end{document}